\documentclass[preprint,12pt,number]{elsarticle}

\usepackage{graphicx}

\usepackage{dcolumn}
\usepackage{bm}
\usepackage{color}

\journal{Physics Letters B}

\begin{document}

\begin{frontmatter}

\title{
  Stochastic Estimation of Nuclear Level Density in the Nuclear Shell Model: 
  An Application to Parity-Dependent Level Density in $^{58}$Ni 
}
\author[cns]{Noritaka Shimizu}
\ead{shimizu@cns.s.u-tokyo.ac.jp}
\author[cns,jaea]{Yutaka Utsuno}
\author[tsukuba]{Yasunori Futamura}
\author[tsukuba,crest]{Tetsuya Sakurai}
\author[senshu]{Takahiro Mizusaki}
\author[cns,phystokyo,leuven]{ Takaharu Otsuka }

\address[cns]{Center for Nuclear Study, 
  the University of Tokyo, 
  Hongo Tokyo 113-0033, Japan}
\address[jaea]{Advanced Science Research Center, 
  Japan Atomic Energy Agency, 
  Tokai, Ibaraki 319-1195, Japan }
\address[tsukuba]{Faculty of Engineering, Information and Systems, 
   University of Tsukuba, Tsukuba, 305-8573, Japan}
\address[crest]{CREST, Japan Science and Technology Agency,
  Kawaguchi, 332-0012, Japan}
\address[senshu]{Institute of Natural Sciences, 
  Senshu University, Tokyo 101-8425, Japan}
\address[phystokyo]{Department of Physics, the University of Tokyo, 
  Hongo Tokyo 113-0033, Japan }
\address[nscl]{National Superconducting Cyclotron Laboratory, 
  Michigan State University, East Lansing, MI 48824, USA}
\address[leuven]{Instituut voor Kern- en Stralingsfysica, 
   Katholieke Universiteit
   Leuven, B-3001 Leuven, Belgium}

\begin{abstract}
    We introduce a novel method to obtain 
    level densities in large-scale shell-model calculations.
    Our method is a stochastic estimation of eigenvalue count based 
    on a shifted Krylov-subspace method, 
    which enables us to obtain level densities 
    of huge Hamiltonian matrices. 
    This framework leads to a successful 
    description of both low-lying spectroscopy 
    and the experimentally observed equilibration of
    $J^\pi=2^+$ and $2^-$ states in $^{58}$Ni in a unified manner. 
\end{abstract}


\end{frontmatter}

\section{Introduction}
\label{sec:intro}

Nuclear level density, which constitutes a very large uncertainty 
in estimating cross sections of compound nuclear reactions, 
plays an essential role in describing those reactions 
usually with the Hauser-Feshbach theory \cite{hauser}.
Systematic data for compound nuclear reactions, including 
$(n,\gamma)$ reaction data, 
are highly needed in many applications including 
nuclear astrophysics and nuclear engineering. 
The $(n,\gamma)$ cross sections for short-lived nuclei, however, 
cannot be directly measured, and therefore 
an accurate estimate of nuclear level density 
is crucial for those needs \cite{ld-astro}. 
While nuclear level density is known to well follow 
phenomenological formulas, such as the backshifted Fermi-gas model 
and the constant temperature model, their model parameters 
depend on nuclear structure and are usually 
derived from experimental data \cite{ld-syst}. 
Moreover, the phenomenological models assume rather simple 
spin-parity dependence of level density, whose validity 
needs to be verified.  
Microscopic theories of nuclear level density 
should thus be developed. 

The large-scale shell-model calculation
provides, in principle, quite reliable level density 
because this method suitably takes two-body correlations into account, 
thus well describing level structures. 
Although the Lanczos diagonalization \cite{lanczos} now 
makes possible the calculation of 
low-lying levels for Hamiltonian 
matrices up to $O(10^{10})$ 
dimensions in an $M$-scheme \cite{caurier_review}, 
its applicability to the level density is much more 
limited because high-lying states are very slow to converge 
in the Lanczos iterations. 
Much effort has been paid to overcoming this limitation
(e.g. the method based on the moment of the matrix
\cite{senkov-zelevinsky, senkov-horoi-ld, johnson-ld}). 
The shell-model Monte Carlo (SMMC) method
has been proposed and developed \cite{smmc-parity,smmc-ni}, 
but it is rather difficult for the SMMC 
to use realistic residual interactions 
which cause the sign problem. 
Recently several realistic interactions have succeeded in 
systematically describing low-lying levels including those of exotic
nuclei \cite{caurier_review, usd-ab, gxpf1b, 
  coraggio_review, otsuka-schwenk, lenzi, sdpf-mu}. 
It is thus quite interesting to investigate whether 
a unified description of low-lying levels and level density 
can be obtained with such modern realistic interactions 
\cite{brown_larsen}.

In this Letter, we apply a novel method of counting eigenvalues 
\cite{futamura-stochastic} to 
the shell-model calculation of level density, 
and demonstrate its feasibility and usefulness.
This method is applicable to any effective interaction
and it straightforwardly provides 
spin-parity dependent level densities. 
More importantly, its computational cost 
is almost at the same order as that of 
the usual Lanczos diagonalization for low-lying states. 
Taking these advantages, 
we are successful in calculating  
spin-parity dependent level densities in $^{58}$Ni 
with a realistic effective interaction, and 
in resolving a puzzle that previous microscopic calculations 
failed to resolve in reproducing no parity dependence of the $2^+$ and 
$2^-$ level densities observed recently \cite{kalymkov2007}. 
The $M$-scheme dimension for the $2^-$ levels 
reaches $1.5\times 10^{10}$, which 
is nearly the current limit of the Lanczos diagonalization.

\section{Formalism}
\label{sec:formalism}

\begin{figure}[htbp]
  \includegraphics[scale=0.45]{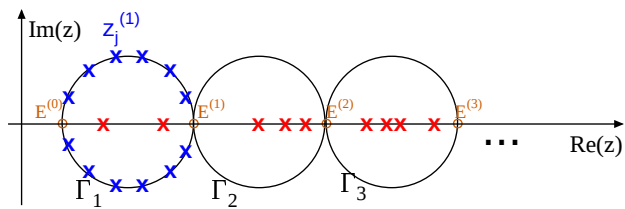}
  \caption{ (Color Online) Schematic view of the contour 
    integral to count the eigenvalues (red crosses).
    Blue crosses denote the discretized mesh points $z_j^{(1)}$ 
    along the integral contour $\Gamma_1$. $E^{(k-1)}$ and $E^{(k)}$ are
    the intersections of $\Gamma_k$ and the real axis.
    \label{fig:contour} }
\end{figure}

We first outline a method for stochastically estimating eigenvalue 
distribution proposed recently  \cite{futamura-stochastic}.
In nuclear shell model calculations, 
a many-body wave function is described as a linear combination 
of a huge amount of many-body configurations, 
which are called the $M$-scheme basis states 
\cite{caurier_review}. 
The shell-model energy is obtained 
as an eigenvalue of the $M$-scheme shell-model Hamiltonian matrix, $H$.
The number of eigenvalues $\mu_k$ in a specified energy range 
$E^{(k-1)}<E<E^{(k)}$ 
is obtained by the contour integral $\Gamma_k$ on the complex plane
in Fig.\ref{fig:contour} as
\begin{eqnarray}
    \label{eq:cinteg}
    \mu_k &=& \frac{1}{2\pi i}
    \oint_{\Gamma_k} dz \ \textrm{tr} ((z-H)^{-1}) \nonumber \\
    &\simeq & \sum_j w_j 
    \textrm{tr}  ((z^{(k)}_j-H)^{-1} ).
\end{eqnarray}
The contour integral on $\Gamma_k$ is numerically obtained by 
discretizing the contour line with mesh points $z^{(k)}_j$
(blue crosses of Fig.\ref{fig:contour}) and 
their weights $w_j$.

The trace of the inverse of matrix in Eq.(\ref{eq:cinteg}) 
is stochastically estimated 
by sampling dozens of random vectors from the whole Hilbert space.
An unbiased estimation is given by 
\begin{equation}
    \label{eq:trace}
    \textrm{tr}( (z-H)^{-1} )
    \simeq 
    \frac{1}{N_s} \sum_s^{N_s} \boldsymbol{v}_s^T (z-H)^{-1} \boldsymbol{v}_s
\end{equation}
where $N_s$ is the number of sample vectors and $\boldsymbol{v}_s$ 
are vectors whose elements take $1$ or $-1$ 
with equal probability randomly
\cite{mctrace}. Note that the norm of $\boldsymbol{v}_s$ 
equals the whole number of states.
Since the Hamiltonian matrix is written in 
the $M$-scheme representation, 
whose basis states are the 
eigenstates of the $z$-component of spin and parity, 
we obtain the total level density with a fixed parity at once.

The remaining task is to obtain 
$\boldsymbol{v}^T_s (z^{(k)}_j-H)^{-1} \boldsymbol{v}_s$, numerically. 
In preceding works, 
the recursion method with the Lanczos algorithm with
was used to compute this value \cite{haydock}, 
but the Lanczos algorithm requires 
reorthogonalization procedure \cite{ss-mizusaki}.
In the present work, 
we adopt the complex orthogonal conjugate gradient 
(COCG) method \cite{takayama-prb}, 
which does not need reorthogonalization.

We briefly explain how to 
calculate the $\boldsymbol{v}^T_s (z^{(k)}_j-H)^{-1} \boldsymbol{v}_s$
with the COCG method.
First, we calculate this value 
at a fixed value, 
$z=z_{\textrm{ref}}$, as a reference.
Instead of calculating the matrix inverse in 
$(z^{(k)}_j-H)^{-1} \boldsymbol{v}_s$,  
we solve a linear equation
$ \boldsymbol{v}_s = (z_{\textrm{ref}}-H) \boldsymbol{x}^{(s)}$ 
with the COCG method for each $s$.
In the COCG method, by omitting the index $s$ for brevity,  the following 
procedure is iterated to solve 
$ \boldsymbol{v}_s = (z_{\textrm{ref}}-H) \boldsymbol{x}^{(s)}$ 
until the residual $\boldsymbol{r}_n$ is small enough: 
\begin{eqnarray}
    \boldsymbol{x}_{n} &=& \boldsymbol{x}_{n-1}
    + \alpha_{n-1} \boldsymbol{p}_{n-1}
    \nonumber \\
    \boldsymbol{r}_{n} &=& \boldsymbol{r}_{n-1}  + \alpha_{n-1}
    (z_{\textrm{ref}}-H) \boldsymbol{p}_{n-1}
    \nonumber \\
    \boldsymbol{p}_{n} &=& \boldsymbol{r}_{n}  + \beta_{n-1} \boldsymbol{p}_{n-1}
    \label{eq:cocg}
\end{eqnarray}
with $\boldsymbol{x}_{0} = \boldsymbol{p}_{-1} = 0$,
$\boldsymbol{r}_{0} = \boldsymbol{v}_s$, 
$\alpha_{n} = \boldsymbol{r}_{n}^T\boldsymbol{r}_{n}/
\boldsymbol{p}_{n}^T(z_{\textrm{ref}}-H) \boldsymbol{p}_{n}$,  
and 
$\beta_{n} = \boldsymbol{r}_{n+1}^T\boldsymbol{r}_{n+1}/
\boldsymbol{r}_{n}^T\boldsymbol{r}_{n}$.
In actual calculations,  $ \boldsymbol{x}^{(s)} $ $(s=1,2,...,N_s)$ 
are obtained simultaneously with a block algorithm (BCOCG) 
\cite{futamura-sbcgrq}
with economical implementation for computing block bilinear 
form \cite{du-futamura-sakurai} for efficient computation.

Second, we solve 
$ \boldsymbol{v}_s = (z^{(k)}_j-H) \boldsymbol{x}^{(sjk)}$ 
for other $z=z^{(k)}_j$.
If we perform the COCG method for each $z^{(k)}_j$ individually, 
an impractically large amount of computation is required. 
However, the shifted Krylov-subspace method 
\cite{shift-krylov, yamamoto-jpsj}
enables us to solve these equations at once.
The basic idea behind this method is 
the invariance property of the shifted Krylov subspace: 
The Krylov subspace of the matrix $z-H$ and vector 
$\boldsymbol{v}$, which is defined as 
\begin{eqnarray}
    && K_n(z-H, \boldsymbol{v}) \nonumber \\
    &=& \textrm{span}\{\boldsymbol{v}, (z-H)\boldsymbol{v}, 
    (z-H)^2\boldsymbol{v}, ..., (z-H)^{n-1}\boldsymbol{v} \}
    \nonumber \\
    &=& \textrm{span}\{\boldsymbol{v}, H\boldsymbol{v}, 
    H^2\boldsymbol{v}, ..., H^{n-1}\boldsymbol{v} \}
    = K_n(H, \boldsymbol{v}), 
    \label{eq:shift-krylov}
\end{eqnarray}
is independent of $z$.
Since the COCG method is one of the Krylov-subspace methods, 
its solution of $z=z_{\textrm{ref}}$ is described in this subspace. 
The solution of any $z$ is also obtained 
in the same subspace simultaneously, if $n$ is large enough.
Thus, by utilizing the COCG solution at $z=z_{\textrm{ref}}$, 
the shifted method allows us to avoid 
the time-consuming calculations of 
matrix-vector product for any $z^{(k)}_j$, 
and to obtain the level density in the whole energy region at once. 
In practice, we adopt the shifted block CG-rQ (SBCG-rQ) method 
\cite{futamura-sbcgrq} for efficient computation.
This computation demands memory usage for only three vector blocks
($\boldsymbol{x}_{n}, \boldsymbol{r}_{n},$ and $\boldsymbol{p}_{n}$)
and a working area for the matrix-vector products, 
while all eigenvectors are stored in the Lanczos method.

For spin-dependent ($J$-dependent) level density, 
we replace $\boldsymbol{v}_s$ by 
$\boldsymbol{v}^J_s = P_J \boldsymbol{v}_s$, 
where $P_J$ is the $J$-projection operator. 
This projection is actually realized by 
filtering out a specified range of the eigenvalue of 
the $J^2$ operator
by the Sakurai-Sugiura method 
\cite{ss-mizusaki, ss-method}. 
Note that the $J$-projection is not 
necessary during the COCG iteration in this work, 
whereas $J$-projection is performed 
at every iteration to suppress numerical errors
in the $J$-projected Lanczos method 
\cite{caurier_review, ss-mizusaki}. 
This is because an undesirable-$J$ component 
caused by numerical error does not affect 
the inner product $(\boldsymbol{v}^J_s)^T \boldsymbol{x}_n^\sigma$. 
Even using such a sophisticated method, 
the eigenvalue count of a huge matrix 
requires high-performance computing. 
We combine the nuclear shell-model code ``KSHELL'' \cite{kshell} 
and the eigenvalue-solver library ``z-Pares'' \cite{zpares}, 
which enable us to utilize the latest supercomputer efficiently.

It is worth noting a pioneering work for the level density 
using the Lanczos method by estimating the traces with 
random vectors in Ref.\cite{grimes}.

\section{Benchmark test}
\label{sec:bench}

\begin{figure}[h]
    \centering
    \includegraphics[scale=0.4]{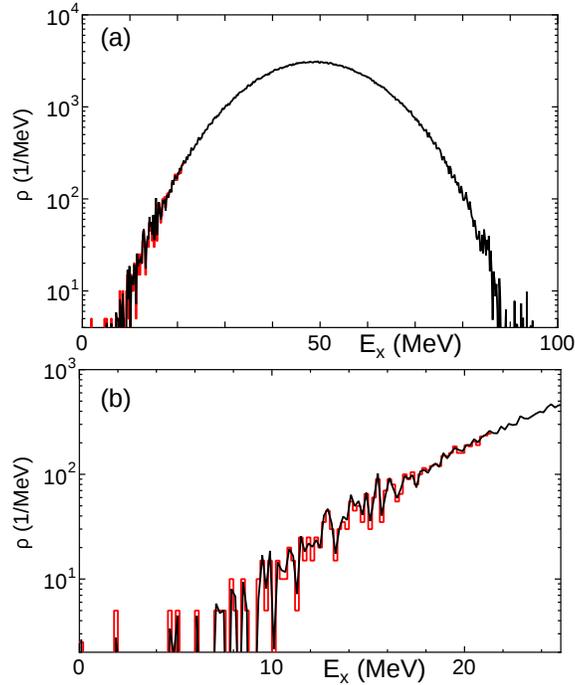}
    \caption{
      (a) Total level density of $^{28}$Si 
      by stochastic estimation (black line) 
      and the  Lanczos method (red solid stair)
      as a function of the excitation energy, $E_x$. 
      (b) Low-energy region of (a). }
    \label{fig:si28-ld}
\end{figure}

We demonstrate how this stochastic estimation works well 
in a small system. 
Figure \ref{fig:si28-ld} shows the level density  of $^{28}$Si 
with the USD interaction \cite{usd}.
The level density is obtained with $N_s=32$ and 
a 200 keV energy bin.
We compare the present estimation with the exact values, 
which are provided by counting the 1000 lowest states
calculated with the Lanczos method.
The present estimation successfully reproduces the exact values
in 8\% error at around $E_x=$ 15~MeV 
with such a small bin.
Non-empirical error evaluation remains as future work 
\cite{trace-error}.

\begin{figure}[h]
    \centering
    \includegraphics[scale=0.4]{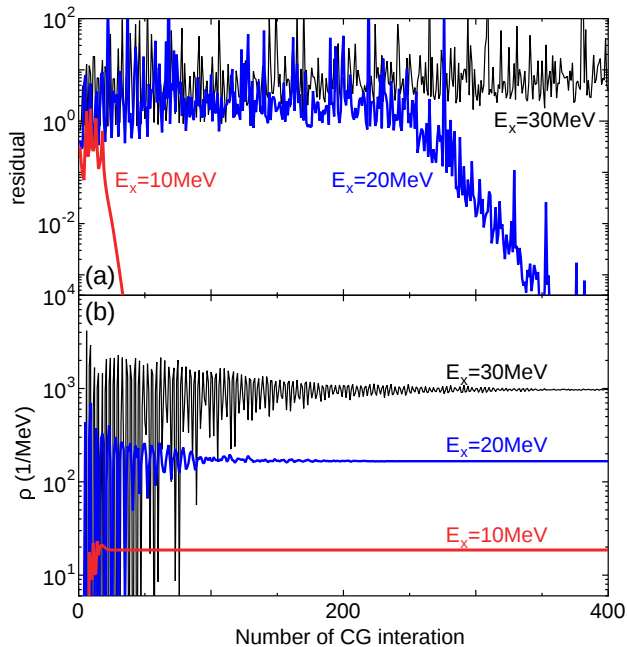}
    \caption{ (Color online)
      Convergence patterns of the COCG and 
      stochastic estimation of the total level density 
      of $^{28}$Si as a function of 
      the number of iterations of the BCOCG method. 
      (a) Residual of the BCOCG iteration $|r_n|/|v_s|$
      at $z=E_{\textrm{gs}}+E_x$ with $E_{\textrm{gs}}$ 
      being the ground-state energy.
      (b) Total level densities at $E_x=$ 10 (red), 
      20 (blue), 30 (black) MeV. }
    \label{fig:si28-conv}
\end{figure}

Figure \ref{fig:si28-conv} (a) shows 
the residual $|r_n|/|v_s|$ of the BCOCG 
against the iteration number in the $^{28}$Si case. 
The residual rapidly converges to zero at low energies
of practical interest.
However, the convergence becomes worse in 
high excited energies as the level density increases.
While a similar tendency is also seen 
for the resultant level densities in Fig.~\ref{fig:si28-conv} (b), 
their convergence is much faster than the residual
and is quite stable. 
For example, only 28 iterations 
are required for the convergence of the level density
at $E_x = $10~MeV.

\section{Parity-equilibration of $^{58}$Ni level density}
\label{sec:58ni}

Now that the feasibility of the present method has been confirmed, 
we can move on to its application. 
Here we will demonstrate that the $2^+$ and $2^-$ level densities 
in $^{58}$Ni \cite{kalymkov2007} are excellently reproduced 
with the present method. 
To describe both parity states 
of nuclei around $^{58}$Ni with the shell model in a practical way, 
we take the $0\hbar\omega$ and $1\hbar\omega$ states 
in the full $sd$+$pf$+$sdg$ valence shell 
for natural- and unnatural-parity states, respectively. 
Since $^{58}$Ni is located in the middle of the $pf$ shell, 
higher $\hbar\omega$ states are expected to be dominant 
in relatively high excitation energies, which will be 
confirmed later. 
The full $1\hbar\omega$ calculation is still impractical for $^{58}$Ni, 
and we truncate the basis states by excluding configurations involving 
more than six-particle excitations from the $0f_{7/2}$ filling
configurations.
We have confirmed that this truncation has a minor effect on 
the level density in the $0\hbar\omega$ calculation.
The resulting $M$-scheme dimension for the $2^-$ states in $^{58}$Ni,  
$1.5\times 10^{10}$, is much beyond the capability of 
the direct counting of eigenstates by the Lanczos diagonalization, 
but within the scope of the present method. 
We remove the spurious center-of-mass contamination associated 
with $1\hbar\omega$ excitation by using  
the Lawson method \cite{lawson}, whereas
this removal is usually not included in SMMC.
The spurious states are lifted up to $E_x \sim600$~MeV,  
thereby separated clearly 
with the Hamiltonian $H'=H+\beta H_{\textrm{CM}}$
with $\beta \hbar\omega/A = 10$~MeV. 

The effective interaction taken in this study is a natural extension 
of the SDPF-MU interaction for the $sd$-$pf$ shell \cite{sdpf-mu} 
to the one for the $sd$-$pf$-$sdg$ shell. Namely, 
exactly in the same way as the
SDPF-MU, the present interaction consists of 
the USD interaction \cite{usd} for the $sd$ shell, 
the GXPF1B interaction \cite{gxpf1b} for the $pf$ shell, 
and a refined monopole-based universal interaction ($V_{\textrm{MU}}$) 
\cite{vmu, utsuno-inpc} for the rest of the two-body matrix elements. 
It is noted that the SDPF-MU interaction well describes 
low-lying $1\hbar\omega$ levels \cite{ni55, si-unnatural} 
as well as $0\hbar\omega$ states \cite{sdpf-mu}. 
Going back to the present $sd$-$pf$-$sdg$-shell Hamiltonian, 
the single-particle energies (SPEs) of the $sdg$ orbits 
are left to be determined,    
while those of the $sd$-$pf$ orbits are 
taken from the SDPF-MU interaction.
Here, the SPEs for the $0g_{9/2}$, $1d_{5/2}$, and 
$2s_{1/2}$ orbits are fixed 
to reproduce experimental spectroscopic strengths 
around $^{58}$Ni, as shown in the next paragraph. 
On the other hand, experimental information on the remaining upper 
$sdg$ orbits is missing near $^{58}$Ni. 
These SPEs are chosen to fit
the empirical neutron effective SPEs on top of  $^{90}$Zr \cite{vmu}.

The $1\hbar\omega$ shell gaps (the $sd$-$pf$ and $pf$-$sdg$ shell gaps)
dominate the overall positions of the $1\hbar\omega$ 
states and thus, their level densities. 
The reliability of those shell gaps in the present Hamiltonian 
is examined by comparing theoretical and experimental
spectroscopic strengths ($C^2S$)
for the one-proton ($\pm 1p$) and one-neutron ($\pm 1n$) transfer 
reactions in $^{58}$Ni. 
The results are summarized in Table~\ref{tab:ll}, 
where the $sd$-$pf$ shell gaps and the $pf$-$sdg$ shell gaps
are probed by the $-1p$ and $-1n$ reactions ($^{57}$Co and $^{57}$Ni)
and the $+1p$ and $+1n$ reactions ($^{59}$Cu and $^{59}$Ni), 
respectively. 
Aside from some difference in $C^2S$ fragmentation 
for the $3/2^+$ levels in $^{57}$Ni 
and the $1/2^+$ levels in $^{59}$Ni,
the present shell-model results  
agree well with the experimental data, thus confirming 
suitable effective SPEs for upper $sd$ ($1s_{1/2}$ and $0d_{3/2}$) 
and lower $sdg$ ($0g_{9/2}$, $1d_{5/2}$ and $2s_{1/2}$) orbits. 
In addition,
the $3^-_1$ level in $^{58}$Ni is also well reproduced:  
4.55~MeV (Cal.) vs. 4.47~MeV (Exp.). 

\begin{table}[t] 
    \begin{tabular}{ccllcll} 
        \hline
        \hline
        Nucl. & $J^\pi$ 
        & \multicolumn{2}{c}{ $E_x$ (MeV)}   
        &  \multicolumn{3}{c}{  $C^2S$ }  \\ \cline{3-4} \cline{5-7}
        & &  Cal.  & Exp. & $j$ & Cal.  & Exp.  \\
        \hline
        $^{57}$Co 
        & $7/2^-_1$ & 0      & 0     
        & $\pi 0f_{7/2}^{-1}$ & 5.28 & 4.27, 5.53  \\
        & $1/2^+_1$ & 3.037 & 2.981 
        & $\pi 1s_{1/2}^{-1}$ & 0.98 & 1.05, 1.31  \\
        & $3/2^+_1$ & 3.565 & 3.560 
        & $\pi 0d_{3/2}^{-1}$ & 1.70 & 1.50, 2.33  \\
        \hline
        $^{57}$Ni 
        & $3/2^-_1$ & 0     & 0     
        & $\nu 1p_{3/2}^{-1}$ & 1.14 & 1.04, 1.25, 0.96  \\
        & $1/2^+_1$ & 5.581 & 5.580 
        & $\nu 1s_{1/2}^{-1}$ & 0.51 & 0.62, 1.08 \\
        & $3/2^+_1$ & 5.579 & 4.372 
        & $\nu 0d_{3/2}^{-1}$ & 0.29 & 0.01 \\
        & $3/2^+_2$ & 6.093 & 6.027 
        & $\nu 0d_{3/2}^{-1}$ & 0.22 & 0.66, 0.54\\
        \hline
        $^{59}$Cu
        & $3/2^-_1$ & 0     & 0     
        & $\pi 1p_{3/2}^{+1}$ & 0.53 & 0.46, 0.49, 0.25  \\
        & $9/2^+_1$ & 3.139 & 3.023 
        & $\pi 0g_{9/2}^{+1}$ & 0.26 & 0.24, 0.32, 0.27  \\
        \hline
        $^{59}$Ni 
        & $3/2^-_1$ & 0     & 0     
        & $\nu 1p_{3/2}^{+1}$ & 0.51 & 0.82, 0.33 \\
        & $9/2^+_1$ & 3.053 & 3.054 
        & $\nu 0g_{9/2}^{+1}$ & 0.63 & 0.84, 0.39 \\
        & $5/2^+_1$ & 4.088 & 3.544 
        & $\nu 1d_{5/2}^{+1}$ & 0.04 & 0.03 \\
        & $5/2^+_2$ & 4.595 & 4.506 
        & $\nu 1d_{5/2}^{+1}$ & 0.30 & 0.23, 0.14 \\
        & $1/2^+_1$ & 4.399 & 5.149 
        & $\nu 2s_{1/2}^{+1}$ & 0.00 & 0.09 \\
        & $1/2^+_2$ & 5.492 & 5.569 
        & $\nu 2s_{1/2}^{+1}$ & 0.18 & 0.02 \\
        & $1/2^+_3$ & 5.589 & 5.692
        & $\nu 2s_{1/2}^{+1}$ & 0.02 & 0.13 \\
        \hline
        \hline
    \end{tabular}
    \caption{ Excitation energies and one-nucleon 
      spectroscopic factors ($C^2S$) of the ground states
      and low-lying unnatural parity states 
      in $^{57}$Ni, $^{59}$Ni, $^{57}$Co, and $^{59}$Cu compared 
      between the shell-model results and experiments \cite{ensdf}.
      The $C^2S$ values are for one-nucleon transfer reactions 
      of $^{58}$Ni, and the transferred nucleon and its orbital are specified 
      in the fifth column. }
    \label{tab:ll}
\end{table}

\begin{figure}[htbp]
  \includegraphics[scale=0.45]{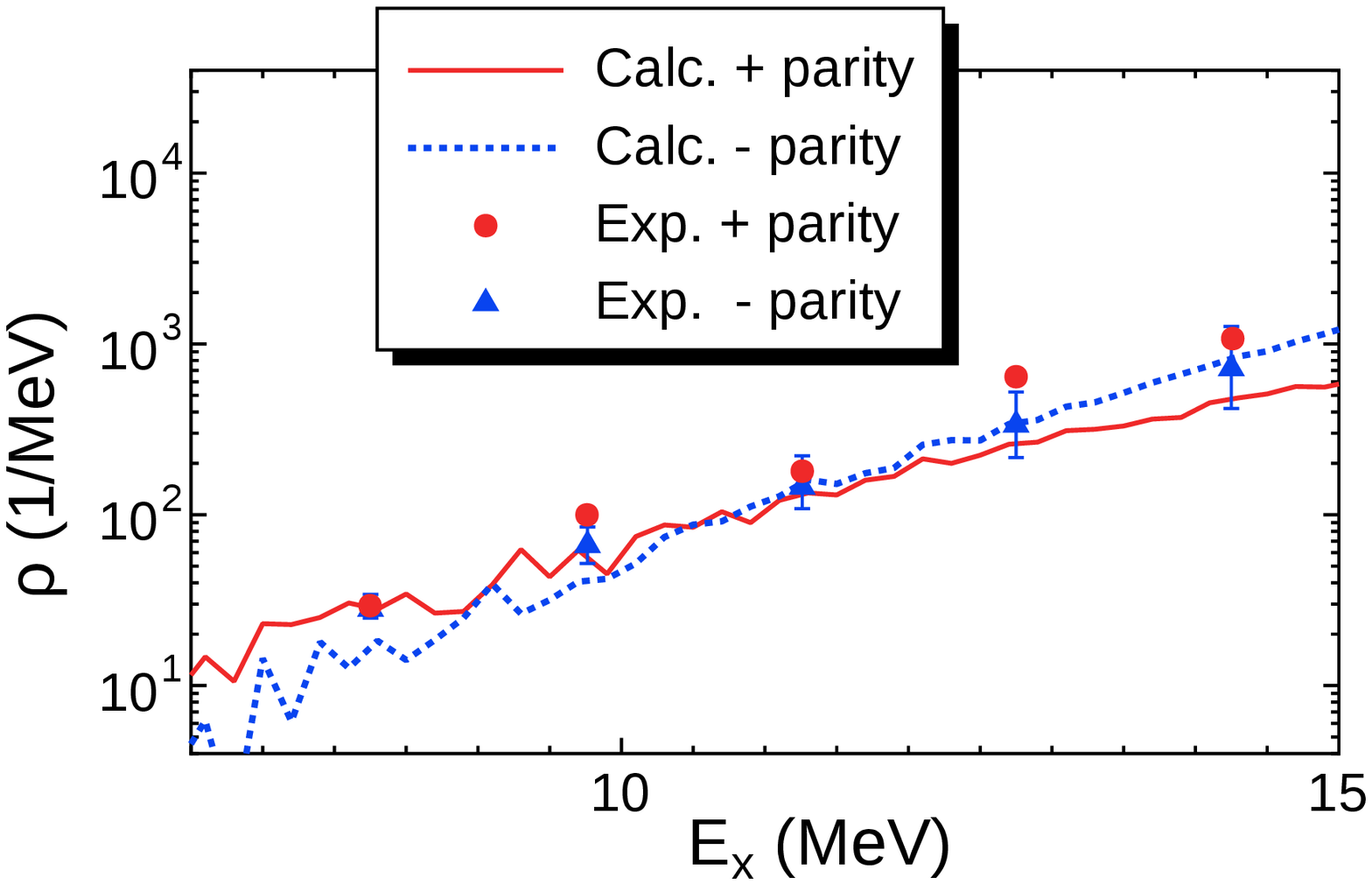}
  \caption{ (Color Online) Level density 
    of of $^{58}$Ni as a function of excitation energy.
    The theoretical results are shown 
    by the red solid ($J^\pi=2^+$) and blue dotted ($2^-$) lines, 
    while the experimental values are shown by 
    circles ($2^+$) and triangles ($2^-$) \cite{kalymkov2007}.
    \label{fig:ratio-ni58} 
  }
\end{figure}

By utilizing this realistic interaction and 
the new stochastic method for estimating level density, 
we can calculate $2^+$ and $2^-$ level densities in $^{58}$Ni. 
The $2^-$ level densities are calculated with $N_s=16$ 
and 1000 BCOCG-iterations which are large 
enough to reach reasonable convergence. 
Figure~\ref{fig:ratio-ni58} compares 
$2^{\pm}$ level densities in $^{58}$Ni
between the experimental data \cite{kalymkov2007} and 
the present calculation. 
What is surprising in this experimental data is that 
the equilibration of $2^+$ and $2^-$ levels 
occurs already at $E_x \sim 8$~MeV at variance with 
SMMC and HFB-based \cite{hfb-ld} estimates in which 
the equilibration is realized only at around 20~MeV \cite{kalymkov2007}. 
Both of the calculations overestimate the $2^+$ level densities 
and underestimate the $2^-$ level densities 
in low excitation energies.
In contrast, the present calculation correctly 
reproduces the early onset of the equilibration 
on the basis of good agreement of the $2^+$ and $2^-$ level 
densities there.
It is worth pointing out that nucleon excitation 
from the $sd$ shell 
accounts for about half of the $2^-$ level 
densities. With increasing excitation energy, 
the present calculation begins to underestimate 
the $2^+$ level densities 
probably because of the absence of $2\hbar\omega$ levels. 
The calculated $2^-$ levels, on the other hand, 
keep following the experimental data up to 15~MeV, because 
the $3\hbar\omega$ level densities are expected to grow at higher
energies.

\section{Summary}
\label{sec:summary}

In summary, we introduced a stochastic estimation of 
the level density in nuclear shell-model calculations
and demonstrated its feasibility.
This method is based on the shifted Krylov-subspace method 
and enables us to obtain the 
level density of a huge-dimension matrix
beyond the limitation of the conventional Lanczos method.
The contamination of spurious center-of-mass excitation 
is clearly removed by the Lawson method. 
By combining proven SDPF-MU and $V_{\textrm{MU}}$ interactions,
we constructed a realistic effective interaction
which successfully describes 
low-lying levels and their spectroscopic factors around $^{58}$Ni.
With this realistic interaction and the stochastic method, 
we obtained the level densities of  $J^\pi=2^+$ and $J^\pi=2^-$ in $^{58}$Ni.
These densities are 
in excellent agreement with the experimental results
and show the equilibration at $E_x \ge 8$~MeV, 
whereas the preceding microscopic calculations 
showed strong parity dependence.
The present framework bridges the low-lying spectroscopy 
and microscopic understanding in statistical region;  
thus, further studies in this direction should be quite promising.
Moreover, the application to {\it ab initio} shell-model calculations
in light nuclei 
is expected.

\section*{Acknowledgments}
\label{sec:ack}

    This work has been supported by 
    the HPCI Strategic Program from MEXT, 
    CREST from JST, 
    the CNS-RIKEN joint project 
    for large-scale nuclear structure calculations, 
    and KAKENHI grants (25870168, 23244049, 15K05094) 
    from JSPS, Japan.
    The numerical calculations were performed 
    on K computer at RIKEN AICS (hp140210, hp150224)
    and COMA supercomputer at the University of Tsukuba.


\end{document}